\documentclass[letterpaper, 10 pt, conference]{ieeeconf}  

\IEEEoverridecommandlockouts                              

\usepackage{graphics}       
\usepackage{epsfig}         
\usepackage{times}          
\usepackage{amsmath}        
\usepackage{amssymb}        
\usepackage{color}
\usepackage{array}
\usepackage{accents}
\usepackage{url}
\usepackage{xcolor}

\usepackage{cite}
\usepackage{amsmath,amssymb,amsfonts}
\usepackage{algorithmic}
\usepackage{graphicx}
\usepackage{textcomp}
\usepackage{xcolor}

\title{\LARGE \bf Design of PI speed controller of motor drive with hardware and software implementation constraints}

\author{Eleonora Brasili, Gianluca Ippoliti, Giuseppe Orlando, Michael Ruderman  
\thanks{E. Brasili, G. Ippoliti, G. Orlando are with Università Politecnica delle Marche, 60131 Ancona, Italy.}%
\thanks{M. Ruderman is with University of Agder, 4879 Grimstad, Norway. \newline Correspondence to: email {\tt\small michael.ruderman@uia.no}}%
}

\begin{document}

\maketitle
\thispagestyle{empty}
\pagestyle{empty}

\bstctlcite{references:BSTcontrol}

\begin{abstract}
Integrated motor drive systems, commercially available in industrial applications, may pose several hardware and software constraints for customized design of an external feedback controller. While the cascade-internal (i.e. current or torque) control loop operates often at higher sampling rates and allows for an automatic tuning by the manufacturer firmware, the outer speed control loop depends largely on the motor drive load characteristics and can have additional restrictions in an customized implementation and tuning. The paper provides a practical case study of designing a PI speed control of the motor drive with ball-screw linear stage application. The unknown system parameters, necessary for a loop shaping, are first identified with a limited set of simple yet well-specified experiments. Afterwards, a robust and performing, at the same time, loop shaping is achieved by taking into account also the reduced sampling rate of a sequential control cycle and saturations of torque controller. Experiments demonstrate that the designed and implemented PI speed controller is expectedly performing over the large range of operation velocities.             
\end{abstract}

\section{Introduction}
\label{sec:1}
Speed regulation of electric motor drives plays a fundamental role in modern industrial automation, enabling accurate motion control and efficient operation of a wide range of machinery and production systems. Its applications span from conventional industrial equipment, such as conveyor belts, pumps, fans, compressors, and lifting systems, to more sophisticated applications, including machine tools, robotic manipulators, and automated manufacturing lines. In these scenarios, accurate speed tracking is essential to ensure productivity, process repeatability, operational safety, and energy efficiency \cite{Li2024}. The increasing integration of electric drives into industrial automation architectures has further emphasized the importance of reliable and high-performance motion control, particularly when operating conditions involve variable mechanical loads, disturbances, and demanding dynamic requirements. Among the available control strategies, proportional-integral controllers remain widely adopted in industrial motor drives due to their relatively simple structure, straightforward implementation, and satisfactory performance in a broad range of applications.

Tuning of PI (or PID) feedback controllers in industrial processes belong to standard control engineering problems, cf. e.g. \cite{cominos2002pid,AstromHagglund2006} and references therein. Still, each particular application and among all the case-specific constraints associated with the accessible process modeling and identification, on the one hand, and with the customized hardware and software of the control unit, on the other hand, require design of the feedback loop to be performed systematically. Only few related works are to be mentioned here (due to space limits) as relevant for the present paper, while the loop-shaping and the resulting different transfer characteristics of the closed-loop, see e.g. \cite{maciejowsk1990multivariable,doyle2013feedback} for basics, are in background of our design. A frequency loop-shaping with integrated system identification was presented for PID controller tuning in \cite{grassi2001integrated}. A practical gain tuning of PI controllers for PMSM drives was proposed in e.g. \cite{tursini2002real}. A cascade of PI speed control and reduced-order PI observer was demonstrated later in \cite{son2015robust}. Moreover, robust speed control of motors and motor-drives can be achieved by coprime factorization (see e.g. \cite{doyle2013feedback}) based two degrees-of-freedom controller design \cite{umeno1991robust}, and so-called disturbance observer compensation techniques \cite{ohishi1987microprocessor}.

The present work provides a practical case study of designing a PI speed control of the motor drive with ball-screw linear stage application. It takes into account the minimal necessary parameters identification, control saturation and time delay constraints, as well as limitations posed by the given system hardware and software. The rest of the paper is organized as follows. A detailed description of the motor drive system is in Section \ref{sec:2}. The identification of the system parameters required for the control design is described in Section \ref{sec:3}, while the control design itself is shown in Section \ref{sec:4}. The experimental control results are demonstrated in Section \ref{sec:5}, and the short conclusions are in Section \ref{sec:6}.

\section{Experimental motor drive system}
\label{sec:2}

\subsection{Hardware setup}
\label{sec:2:sub:1}

\begin{figure}[!h]
\centering \includegraphics[width=0.9\columnwidth]{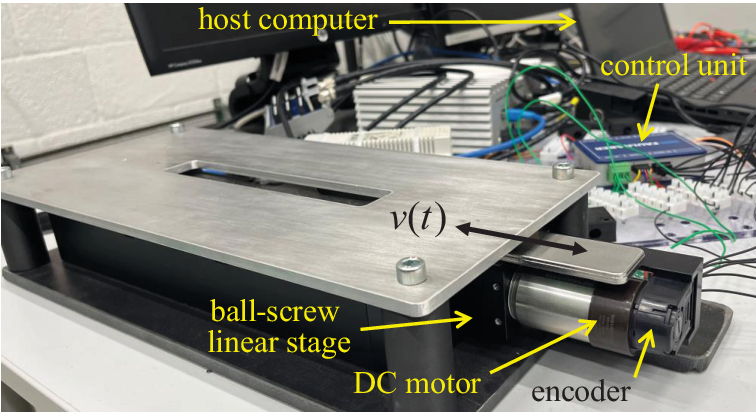}
\caption{Motor driven linear-motion stage system (laboratory, UiA).}
\label{Fig:ExpSetup} 
\end{figure}

The experimental setup, shown in Fig. \ref{Fig:ExpSetup}, consists of a motor-driven linear positioning stage equipped with a ball-screw transmission. The rotational motion of a brushed DC motor is converted into the translational motion of a carriage along a linear axis, in the relative velocity coordinates $v$. 

The motor is (stiffly) mechanically coupled to the ball screw. The mechanical load, seen by the motor, includes the inertia of the rotating components and the equivalent inertia of the translating mass, reflected to the motor shaft through the ball-screw transmission. The system is also affected by the viscous and Coulomb friction, which are considered in the parameter identification procedure described in Section \ref{sec:3}. 
\begin{table}[!h]
\renewcommand{\arraystretch}{1.4}
\centering
\caption{Experimental setup components}
\label{tab:hardware}
\begin{tabular}{lll}
\hline
\textbf{Component} & \textbf{Manufacturer} & \textbf{Model} \\
\hline
DC motor          & FAULHABER & 3557 K 024 CR \\
Linear stage      & Physik Instrumente (PI) & M-414.2PD \\
Motion controller & FAULHABER & MC5005 \\
Software          & FAULHABER & Motion Manager \\
\hline
\end{tabular}
\end{table}
An incremental rotary encoder provides the motor shaft position, from which the rotational velocity is obtained. The encoder signals are processed by the motion control unit, which also provides the electrical power required to drive the motor. The firmware motion controller is connected with a host computer used for system configuration, controller programming, and experimental data acquisition.
The main hardware components and their relevant technical specifications are summarized in Tables \ref{tab:hardware}, \ref{tab:specifications}, respectively \cite{Faulhaber3557,PI_M414}.
\begin{table}[!h]
\renewcommand{\arraystretch}{1.4}
\centering
\caption{Main technical specifications}
\label{tab:specifications}
\resizebox{\columnwidth}{!}{%
\begin{tabular}{llcl}
\hline
\textbf{Component} & \textbf{Parameter} &
\textbf{Symbol} & \textbf{Value} \\
\hline
DC motor
    & Nominal voltage & $U_N$ & 24 V \\
    & Torque constant & $K_t$ & $42.9 \times 10^{-3}$ Nm/A \\
    & Terminal resistance & $R$ & 2.0 $\Omega$ \\
    & Rotor inductance & $L$ & $270 \times 10^{-6}$ H \\
    & Rotor inertia & $J_m$ &
      $6.4 \times 10^{-6}$ kg\,m$^2$ \\
\hline
Linear stage
    & Travel range & -- & 200 mm \\
    & Ball-screw pitch & $p$ & 2 mm/rev \\
    & Maximum velocity & -- & 100 mm/s \\
    & Encoder resolution & -- & 4000 counts/rev \\
    & Design resolution & -- & 0.5 $\mu$m \\
\hline
\end{tabular}%
}
\end{table}
Additional details about the mechanical assembly of the linear stage system, in particularly its kinetic friction, can be found in \cite{ruderman2023robust}.

\subsection{Motion control unit}
\label{sec:2:sub:2}

The motion control unit is MC5005 manufactured by FAULHABER, cf. Fig. \ref{Fig:ExpSetup}, which compactly integrates a power stage, motor feedback acquisition, and cascaded position, velocity, and current control loops \cite{Faulhaber2025}. In the present application, the built-in position and velocity controllers are bypassed purposefully. The velocity regulation is performed by a developed customized PI controller, while the internal current control loop remains active and is low-level implemented in the firmware. Note that the sampling rate of the inner current controller is significantly higher (by about two orders of magnitude) comparing with the developed outer velocity feedback controller, cf. section \ref{sec:2:sub:3}. 

The internal current controller is a PI regulator implemented in the MC5005 firmware. The velocity controller provides a torque demand, which is converted by the drive into a corresponding current reference. The internal PI regulator uses the measured motor current and provides the terminal motor  voltage as output. Its proportional and integral gains, configured through the FAULHABER Motion Manager Software, are $K_{p,I}=0.896$ V/A and $K_{i,I}=5743.59$ V/(A\,s), respectively. It should be emphasized that the regulated motor current is limited by the set parameter value of 1.95 A. The induced electromagnetic torque is expressed as $K_t u(t)$, where $K_t$ is the nominal motor torque constant and $u(t)$ is the motor current in [A] units. For the subsequent control design, the current-loop dynamics are assumed to be sufficiently fast to be neglected, and $u(t)$ is considered as control input to the mechanical system. 

The motor angular velocity $v(t)$ is obtained internally from the incremental encoder and used then by the customized PI controller to calculate the control error. The measured motor current and velocity are also recorded within the MC5005 unit, and can then be read out by the host computer, for purposes of experimental performance evaluation and parameter identification as well. The embedded BASIC programming environment allows the controller to be executed directly on the drive, while limiting implementation of the control logic to the main algebraic operations and integer numbers only.

\subsection{Sequential control program implementation}
\label{sec:2:sub:3}

The customized PI velocity controller is implemented directly within the motion control unit using its built-in BASIC programming environment \cite{Faulhaber2024}. 
The controller operates with a nominal sampling period of $2$ ms, corresponding to an update frequency of $500$ Hz. Unlike the internal current control loop, the customized velocity controller is executed sequentially, with a limited update rate.
At each iteration, the actual motor velocity is acquired and compared with the reference velocity to calculate the tracking error. The proportional and integral contributions are then evaluated to generate the torque command. The integral term is computed using the trapezoidal integration method. Appropriate numerical scaling is applied to perform the calculations using integer arithmetic. 
The resulting control command is limited to the admissible torque range before being applied to the internal current control loop. The sequential program also allows the acquisition of motor velocity, control signals, and timing information for subsequent experimental analysis.

\section{Parameters identification}
\label{sec:3}

The system dynamics described by the lumped-parameters in the relative velocity coordinates $v$ is given by
\begin{equation}\label{eq:3:1}
J \dot{v}(t) + B v(t) + T_c \textrm{sign}\bigl(v(t)\bigr) = K_t u(t),
\end{equation}
where the inertial, damping, and Coulomb friction parameters $J,\, B, \, T_c > 0$
are unknown and need first to be identified. The motor torque constant $K_t$ is given by the manufacturer data-sheet  \cite{Faulhaber3557}. The input channel approximating the low-level regulated motor current is given by $u(t)$. Note that in the following, all system states and parameters are denoted as reflected to the motor side, i.e. in angular motion units and without taking into account the gearing ratio of the ball-screw transmission, this for the sake of simplicity.

The friction parameters are first identified through steady-state experiments performed at different positive and negative motor velocities. For each experiment, the average motor velocity and current are measured. The corresponding electromagnetic torque is obtained from the measured current using the motor torque constant. To account for the observed asymmetry between the two motion directions, the Coulomb friction coefficient $T_c$ in \eqref{eq:3:1} is considered direction-dependent and defined as
\begin{equation}
T_c(v)=
\begin{cases}
T_{c}^{+}, & v>0,\\
T_{c}^{-}, & v<0,
\end{cases}
\end{equation}
respectively. The overall viscous friction coefficient $B$ and the
direction-dependent Coulomb friction coefficients $T_{c}^{+}$ and $T_{c}^{-}$ are then estimated by linear regression of the experimental torque--velocity data. The experimental data and the fitted friction model are compared in Fig.~2. The identified friction parameters are $B=1.972\times10^{-4}$ Nm\,s/rad and $(T_{c}^{+},T_{c}^{-})= (9.794,\,7.114)\times10^{-3}$ Nm.
\begin{figure}[!h]
    \centering
    \includegraphics[width=\columnwidth]{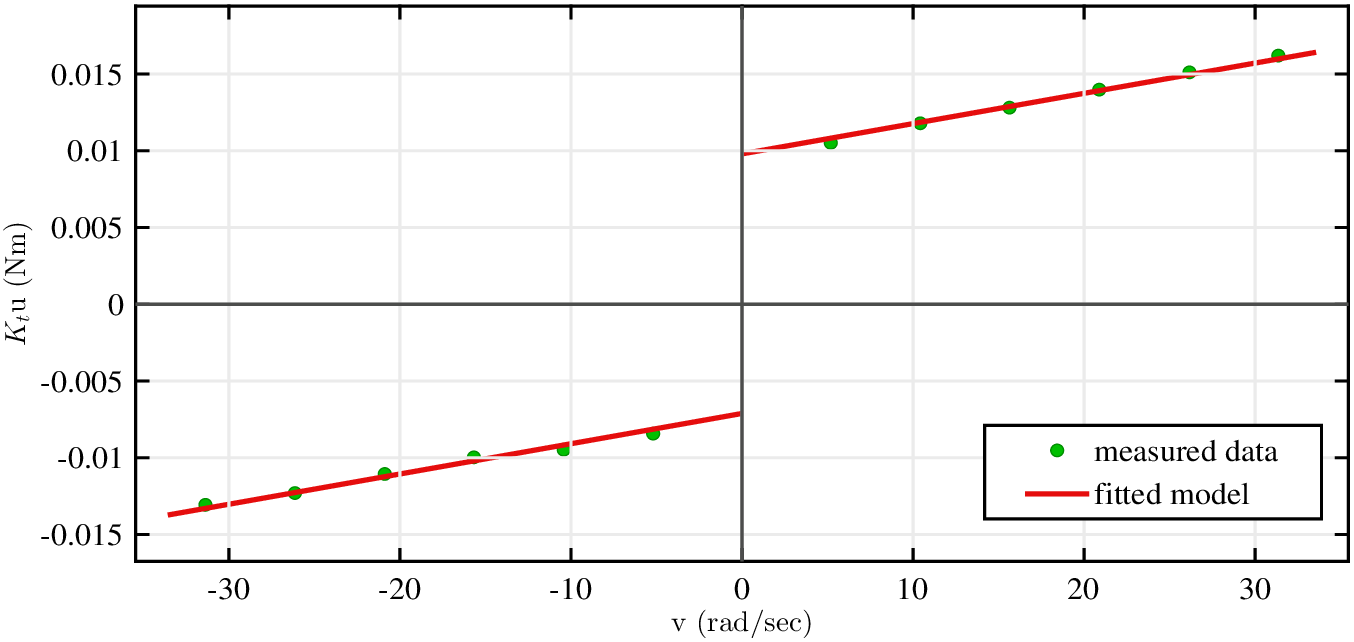}
    \caption{Friction identification: the determined from measurements electromagnetic torque versus motor angular velocity and fitted friction model.}
    \label{fig:friction_identification}
\end{figure}
Note that the identified friction model is not explicitly used in the following linear control design. At the same time, the determined Coulomb friction provides essential information to which extent a constant disturbance will appear during an unidirectional motion with $|v(t)| > 0$, cf. section \ref{sec:4}.

\begin{figure}[!h]
\centering \includegraphics[width=0.98\columnwidth]{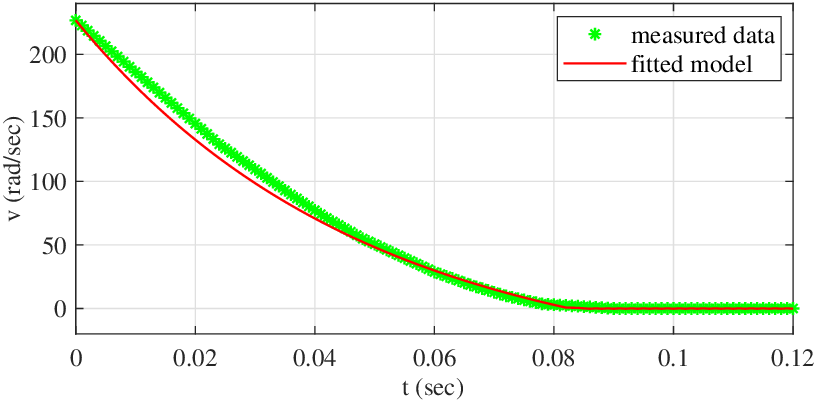}
\caption{Run-out experiment for identifying the total (lumped) system inertia $J$: measured velocity versus the fitted model, cf. \eqref{eq:3:1}.}
\label{Fig:IdentInertia} 
\end{figure}

Once the friction parameters are determined, the total inertia $J$ is estimated through a free-deceleration (run-out) experiment. The motor is initially accelerated to a prescribed (possibly large) velocity, after which the motor drive is disabled. This allows the motor to decelerate freely, without active current regulation, under the effect of mechanical friction. Note that this purposeful experimental scenario is captured by the homogenous (i.e. excitation-free) part of \eqref{eq:3:1}, with the initial condition $v(0) = \textrm{const} \neq 0$. The inertia is then determined by fitting the model response to the measured velocity trajectory, using the previously identified friction coefficients. The comparison between the experimental response and the fitted model is shown in Fig. \ref{Fig:IdentInertia}. The identified total inertia is $J = 9.5 \times 10^{-6}$ kg m$^2$, including both the motor and the equivalent load inertia.

Further we emphasize that the hardware constraints of the linear stage system and software constraints of the control software implementation environment do not allow to perform broadband excitation experiments and corresponding measurement of the system frequency response function. Therefore, the designed above experiments represent a minimal necessary and fully sufficient, at the same time, set of identification steps based on which the loop shaping of velocity feedback control can be performed.

\section{Feedback control design}
\label{sec:4}

The feedback control design follows the principles of loop shaping, cf. e.g. \cite{maciejowsk1990multivariable,doyle2013feedback}, while additional constraints are taken into account as set out below. With respect to a non-negligible time delay of the sequential program cycle inside of the motion control unit, the system input-output transfer behavior is sufficiently approximated by $G(s) = v(s)/u(s)$, cf. \eqref{eq:3:1}, resulting in 
\begin{equation}\label{eq:4:1}
v(s) = \frac{K_m \exp(-\tau s)}{Js + B}\Bigl(u(s) + d \Bigr).
\end{equation}  
Note that comparing to the effective $\tau = 0.001$ sec, an additional closed-loop dynamic of the integrated current controller of electrical motor circuit is neglected in \eqref{eq:4:1}, being secondary for the following loop shaping design. The nonlinear Coulomb friction, cf. \eqref{eq:3:1}, is substituted in \eqref{eq:4:1} by an equivalent disturbance $d$ that is largely valid for unidirectional motion, and it fails to be $d=\textrm{const}$ only during the motion reversals, cf. \cite{ruderman2023book} for details.

\begin{figure}[!h]
\centering 
\includegraphics[width=0.98\columnwidth]{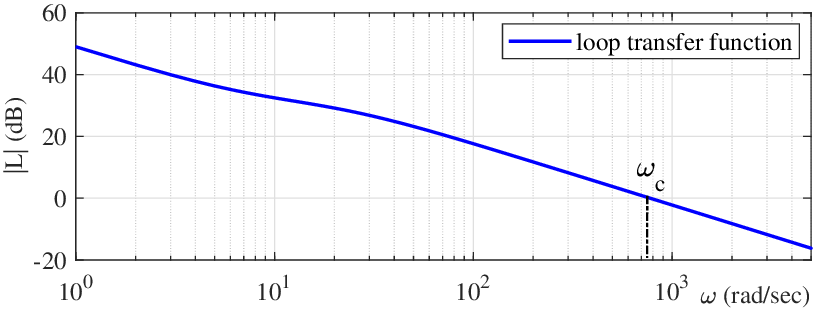}\\[3mm]
\includegraphics[width=0.98\columnwidth]{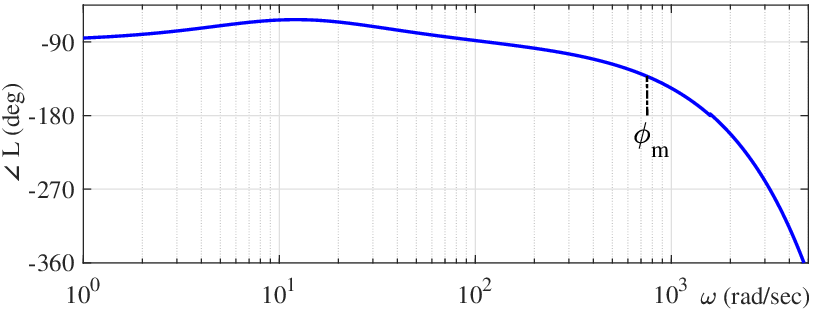}
\caption{Loop transfer function $L(j\omega)$ of designed control system.}
\label{Fig:Loop} 
\end{figure}

\begin{figure}[!h]
\centering 
\includegraphics[width=0.98\columnwidth]{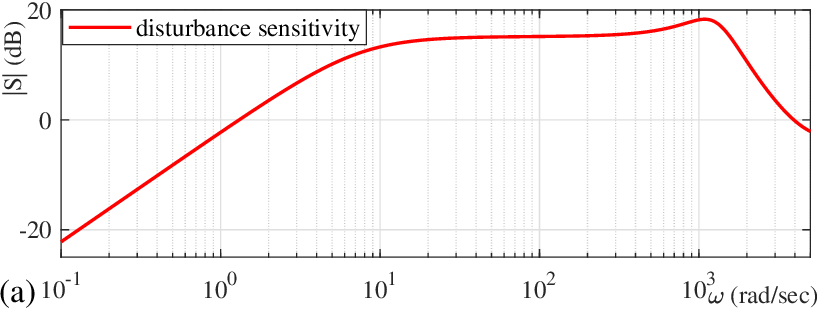}\\[3mm]
\includegraphics[width=0.98\columnwidth]{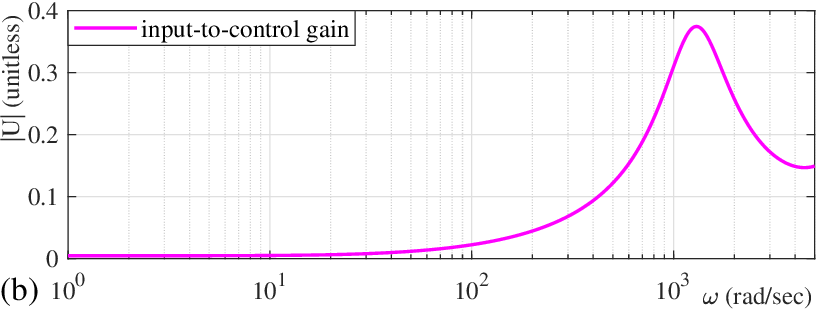}
\caption{Frequency-domain properties of the designed control system: disturbance sensitivity function (a), input-to-control gain (b).}
\label{Fig:Prop} 
\end{figure}

Since the loop delay drastically reduces the phase advance characteristics, see e.g. \cite{mirkin2005control} for basics, of the open-loop transfer function $L(s)=C(s)G(s)$, cf. Fig. \ref{Fig:Loop}, 
the gains of controller $C(s)$ are subject to a conservative assignment, so as to guarantee 
a sufficient phase margin $\phi_m$. Two other relevant transfer characteristics to be taken into account are the disturbance sensitivity function and the input-to-control transfer function 
\begin{equation}\label{eq:4:2} 
S(s) = \frac{v(s)}{d(s)} = \frac{G(s)}{1+L(s)}, \quad U(s) =\frac{u(s)}{v_{r}(s)} = \frac{C(s)}{1+L(s)},
\end{equation} 
respectively. The first one, see Fig. \ref{Fig:Prop} (a), indicates in which frequency range the matched disturbances will be suppressed through the designed feedback loop.
Since an equivalent disturbance representing the Coulomb friction at unidirectional motion is $d=\textrm{const}$, cf. \cite{ruderman2023book}, it can be recognized that $\bigl|S(j\omega)\bigr| \rightarrow 0$ (correspondingly $\rightarrow - \infty$ dB) for $\omega \rightarrow 0$ of the input  $d$. The gain characteristics $\bigl|U(j\omega)\bigr|$ reflects how an external input $w$ injected into the closed loop will affect the control value $u$. Note that apart from the $v_r$-value, cf. \eqref{eq:4:2}, which has $s^{-1}$ characteristics for constant-reference (i.e. step), the $w$ captures in a same way also the sensor measurement noise in feedback. The latter has, however, much lower amplitudes and much higher frequencies than the reference $v_r$. Therefore, based on the $U$-transfer function, the following control system characteristics can be evaluated      
\begin{eqnarray}
\label{eq:4:3}
  \| u \|_{\infty} &=& \| U \|_{1}\, \| v_r \|_{\infty},  \\
  \emph{pow}(u) &=& \| U \|_{\infty}\, \emph{pow}(w),  
\label{eq:4:4} 
\end{eqnarray}
based on the system's and signals norms, cf. \cite{doyle2013feedback} for details. Here, the senor noise power and control signal power are denoted by $\emph{pow}(\cdot)$, respectively, and the infinity norm of the reference and control values are the corresponding least upper bounds of their  absolute value, i.e. $\| \star \|_{\infty} = \sup |\star|$. Recall that the $L_1$ system norm $\| U \|_{1}$ in \eqref{eq:4:3}, also denoted as the system gain, is given by 
\begin{equation}\label{eq:4:4:add}
\| U \|_{1} = \int \limits_0^{\infty} \bigl| \hat{u}(t) \bigr| dt,
\end{equation}
where $\hat{u}(t)$ is the inverse Laplace transform of $U(s)$, i.e. its impulse response. Note that while \eqref{eq:4:4} allows to assess the noise power propagation into the control signal, the equation \eqref{eq:4:3} allows directly to asses and, this way, to constrain the control amplitude in case of the actuator's saturation. This becomes obvious since in case of the step references of the maximal amplitude $V_r$, the corresponding $\| v_r \|_{\infty} = V_r$ while $\| U \|_{1} = \textrm{const} < \infty$, cf. Fig. \ref{Fig:Prop} (b).

The designed with respect to $\tau$, $\bigl|S(j\omega)\bigr|$, and $\| U \|_{1}$, as highlighted above, and evaluated experimentally (see section \ref{sec:5}) PI velocity control is given by
\begin{equation}\label{eq:4:5}
C(s) = \frac{0.1716 s + 1.287}{s}.
\end{equation}
The achieved phase margin of $L(j\omega)$ is $\phi_m = 46.6$ deg, cf. Fig. \ref{Fig:Loop}, which corresponds to about 20\% overshoot in transient step response of the dominant pole pair of the closed-loop system. Note that effectively, the overshoot falls to lower extent, being additionally damped by the nonlinear Coulomb friction, cf. experimental results in Section \ref{sec:5}.

\section{Experimental results}
\label{sec:5}

\begin{figure}[!h]
\centering \includegraphics[width=0.98\columnwidth]{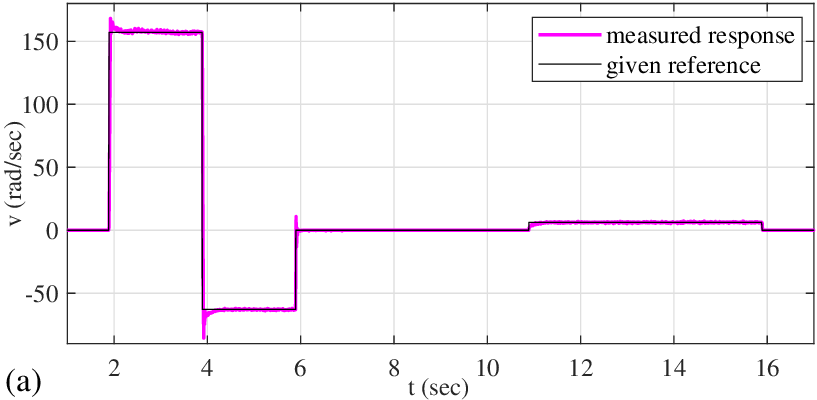}\\[3mm]
\centering \includegraphics[width=0.98\columnwidth]{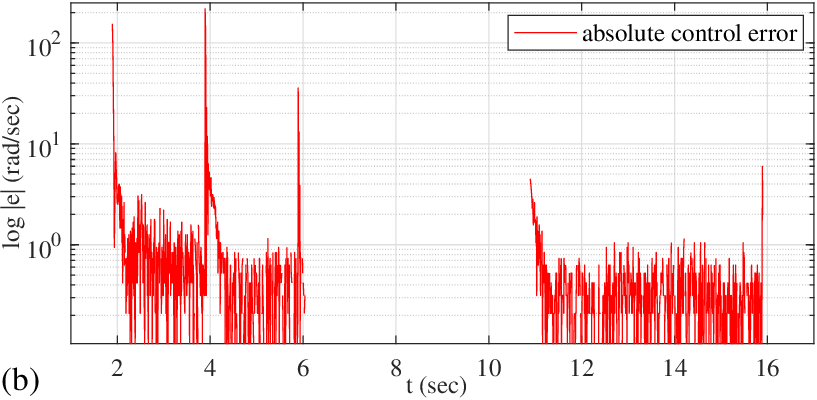}\\[3mm]
\centering \includegraphics[width=0.98\columnwidth]{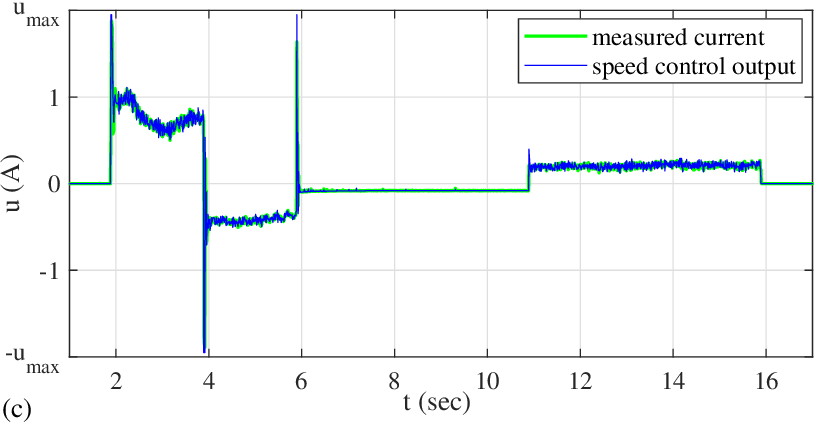}\\
\caption{Experimental results of PI speed control: reference versus measured response (a), absolute control error logarithmical (b), control signal (c).}
\label{Fig:ExpResults} 
\end{figure}

The performance of the implemented PI velocity controller is evaluated experimentally using the reference trajectory with different velocity amplitudes and motion directions. The control error is $e = v_{r}-v$, where $v_{r}$ is the reference velocity. The experimental results are shown in Fig. \ref{Fig:ExpResults}. As can be seen in Fig. \ref{Fig:ExpResults} (a), the measured motor velocity closely follows the prescribed reference, with fast  transient responses and well limited overshoot. Good tracking is achieved for both positive and negative velocities, including motion reversal and low-speed operation. During the constant reference intervals, the velocity remains close to the desired value, with only small fluctuations. The absolute tracking error is shown in Fig. \ref{Fig:ExpResults} (b) on a logarithmic scale, allowing both the large transient errors and the much smaller steady-state errors to be comprehensively observed in the same plot. The error decreases rapidly after each reference change, confirming the fast response of the controller, while the patterns of remaining steady-state error clearly shows the level of the sensor (correspondingly quantization and differentiation) noise. Figure \ref{Fig:ExpResults} (c) shows the speed controller output and the corresponding measured motor current. The control effort changes rapidly during acceleration and motion reversals, while remaining approximately constant once the reference velocity is reached. Although the torque demand briefly reaches its limits during some transients, no sustained saturation is observed for the maximal admissible $|u|_{\max} = 1.95$ A. The measured motor current remains also below its prescribed limits throughout the experiment.

The performance of the implemented PI velocity controller is additionally evaluated, by the error metrics, in terms of the steady-state mean absolute error (MAE), settling time, and overshoot. Four reference steps are considered for the performance evaluation. The final deceleration, visible at the end of Fig. \ref{Fig:ExpResults} (a), is excluded from the analysis, as it occurs after the control action has been disabled and the control loop has been terminated.

For each step $j$, the steady-state MAE is computed over the final $1$ sec of the corresponding constant-reference interval, using the unfiltered measured velocity. Denoting by $\mathcal{W}_j$ the set of samples within this window and by $N_j$ their number, this metric is calculated as
\begin{equation}
\mathrm{MAE}_j =
\frac{1}{N_j}
\sum_{k\in\mathcal{W}_j}
\left|v_{r,j}-v(t_k)\right|,
\end{equation}
where $v_{r,j}$ and $v(t_k)$ denote the reference and measured velocities, respectively, both expressed in rad/s. The mean steady-state MAE, which is characteristic for the control performance over the large range of operation velocities, is then calculated as
\begin{equation}
 \overline{\mathrm{MAE}}_{\mathrm{ss}}
= \frac{1}{4}\sum_{j=1}^{4}\mathrm{MAE}_j.
\end{equation}

The settling time is defined as the time elapsed from each reference change until the motor velocity enters and remains within a prescribed tolerance band. To reduce the influence of isolated velocity fluctuations, a $20$ ms moving-median filter is applied exclusively for settling-time evaluation. Denoting the filtered velocity by $\tilde{v}(t)$, the tolerance band (of 5 \%) is then defined as
\begin{equation}
\delta_j =
\max\left(0.05|\Delta v_j|,
\frac{\pi}{3}\,\mathrm{rad/s}\right),
\end{equation}
where $\Delta v_j$ is the reference step amplitude, expressed in rad/s. The minimum tolerance of
$\frac{\pi}{3}$ rad/s is introduced to reduce the sensitivity of the settling-time evaluation to small velocity fluctuations, particularly at low reference speeds. Denoting by $\mathcal{I}_j$ the set of the samples belonging to the step $j$, the settling time is calculated as
\begin{equation}
t_{s,j}=\min\Bigl\{t_k-t_{0,j}:
\bigl| \tilde v(t_m)-v_{r,j}\bigr|\leq\delta_j, 
\forall m\in\mathcal{I}_j, m\geq k\Bigr\}
\end{equation}
where $t_{0,j}$ denotes the reference change instant. The above definition requires the filtered velocity to remain within the tolerance band until the next reference change.

The percentage overshoot is calculated using the unfiltered measured velocity and normalized by the absolute reference step amplitude:
\begin{equation}
O_j =
\frac{100}{|\Delta v_j|}
\max\left\{0,\,
\max_{t_k\in\mathcal{I}_j}
\Bigl[
\operatorname{sign}(\Delta v_j)
(v(t_k)-v_{r,j})
\Bigr]\right\},
\end{equation}
where $\mathcal{I}_j$ denotes the samples belonging to the step $j$. This definition accounts for both positive and negative reference changes. The minimum and maximum overshoot metrics are determined as $O_{\min}=\min_j O_j$ and $O_{\max}=\max_j O_j$, respectively. The resulting performance indices (correspondingly metrics) are summarized in Table~\ref{tab:PI_performance}.
\begin{table}[!h]
\renewcommand{\arraystretch}{1.4}
\centering
\caption{Experimental performance of the
velocity controller.}
\label{tab:PI_performance}
\footnotesize
\setlength{\tabcolsep}{3pt}
\begin{tabular}{lcc}
\hline
\textbf{Performance index} &
\textbf{Symbol} & \textbf{Value} \\
\hline
Mean steady-state MAE
    & $\overline{\mathrm{MAE}}_{\mathrm{ss}}$
    & 0.309 rad/s \\
Settling time, step 1
    & $t_{s,1}$ & 0.090 s \\
Settling time, step 2
    & $t_{s,2}$ & 0.048 s \\
Settling time, step 3
    & $t_{s,3}$ & 0.024 s \\
Settling time, step 4
    & $t_{s,4}$ & 0.210 s \\
Minimum overshoot
    & $O_{\min}$ & 7.1\% \\
Maximum overshoot
    & $O_{\max}$ & 18.3\% \\
\hline
\end{tabular}
\end{table}

The performance indices reported in Table \ref{tab:PI_performance} are consistent with the experimental responses in Fig. \ref{Fig:ExpResults}. The mean steady-state MAE is $0.309$ rad/s,
indicating a small tracking error during a constant-reference operation. The settling times indicate a fast response to the reference changes, with the longest settling time of $0.210$ sec occurring for the smallest velocity step. This is well explainable by a larger impact of the nonlinear Coulomb friction, see \cite{ruderman2023book} for details. At low velocities, residual fluctuations have a greater influence on the settling-time evaluation. The measured overshoot ranges from $7.1\%$ to $18.3\%$, showing a relatively limited overshoot for all the considered reference changes, which is also in line with loop shaping of the feedback velocity control, cf. section \ref{sec:4}. The fast transient responses remain also well damped.

\section{Conclusions}
\label{sec:6}

This paper presented a practical approach to the design and implementation of a PI speed controller for an electric motor drive, considering the hardware and software constraints of a commercially available motion control platform and its API (Application Programming Interface). The proposed procedure relies on a preliminary experimental identification of the main mechanical parameters using a limited set of relatively simple experiments. These parameters were subsequently employed to perform a systematic frequency-domain loop-shaping design, explicitly accounting for the execution delay associated with the sequential control program and the saturation limits of the internal current control loop. The experimental results obtained on a motor-driven ball-screw linear stage demonstrated the effectiveness of the proposed design procedure over a wide range of operating velocities and motion directions.

\section*{Acknowledgement}
The authors are grateful to Harald Sauvik (UiA) for laboratory help in instrumenting the electronic control unit.

\bibliographystyle{IEEEtran}
\bibliography{references.bib}

\end{document}